\documentclass[12pt,a4paper,final]{iopart}
\usepackage{iopams}  
\usepackage{graphicx}
\usepackage[breaklinks=true,colorlinks=true,linkcolor=blue,urlcolor=blue,citecolor=blue]{hyperref}
\usepackage[]{color}
\usepackage{multirow} 
\usepackage{soul} 
\usepackage{cite} 
\usepackage{pbox}

\begin{document}

\title{Varied reasoning schema in students' written solutions}

\author{Nandana Weliweriya}
\address{Department of Physics, Kansas State University, Manhattan, KS 66506, USA}
\ead{nandee122@ksu.edu}

\author{Justyna P. Zwolak}
\address{STEM Transformation Institute, Florida International University, Miami, FL 33199, USA}
\address{Department of Teaching and Learning, Florida International University, Miami, FL 33199, USA}
\ead{j.p.zwolak@gmail.com}

\author{Eleanor C. Sayre}
\address{Department of Physics, Kansas State University, Manhattan, KS 66506, USA}
\ead{esayre@ksu.edu}

\author{Dean Zollman}
\address{Center for Research and Innovation in STEM Education, Kansas State University, Manhattan, KS 66506, USA}  
\address{Department of Physics, Kansas State University, Manhattan, KS 66506, USA}
\ead{dzollman@phys.ksu.edu}

\begin{abstract}
The Mathematization project investigates students' use of mathematical tools across the undergraduate physics curriculum. As a part of this project, we look at intermediate mechanics students' written homework solutions to understand how they use these tools in approaching traditional mechanics problems. We use a modified version of the ACER (Activation-Construction-Execution-Reflection) framework to analyze students' solutions and to identify patterns of mathematical skills used on traditional problems. We apply techniques borrowed from network analysis and the Resources Framework to build a ``fingerprint'' of students' mathematical tool use. In this paper, we present preliminary findings on patterns that we identified in students' problem solving.
\end{abstract}

\vspace{2pc}
\noindent{\it Keywords}: Mathematization, Undergraduate homework, Problem solving, Network analysis

\section{Introduction}

Homework is a key element in every undergraduate-level physics course for practicing problem solving. Students often have difficulties combining physics ideas with mathematical calculations and selecting among known mathematical tools. Therefore, it is important to explore how students employ mathematics when approaching physics homework problems. 
	
In this paper, we focus on the mathematical tools required in upper-division physics courses, such as evaluating integrals, using differential equations and approximations, or choosing an appropriate coordinate system. Although these techniques are covered in mathematics classes, where students can successfully solve problems, many students still struggle to apply these mathematical tools to problems in physics. This is true especially at the upper-division when problems are more complicated and mathematics more tightly entwined with problem solving~\cite{Wallace10,Pepper12}. In our study, we use the ACER (Activation-Construction-Execution-Reflection) framework~\cite{Wilcox12,Wilcox13} combined with network analytic methods~\cite{Kowalczyk} to investigate how upper-division students' written work shows their mathematical tool use on typical homework problems.
	
The ACER framework was developed using think-aloud interview data on conceptually-rich research-based problems. These problems are different in character from the more traditional problems in our study. Our data are wholly written accounts with sparse reasoning evidence.  However, because students' written homework solutions are common at the upper-division level, they may represent a ubiquitous and easy-to-acquire data source for education research.  Within these constraints, this paper explores how, if at all, the modified ACER framework can be used to analyze written evidence of typical problem solving in a traditional course.  We investigate three kinds of differences: differences in solution paths within one problem; differences among multiple problems of the same type; and differences in students' initial problem solving steps as a function of kind of problem. 

\section{Context}

Our data are drawn from an undergraduate course in Classical Mechanics taught at a large land-grant university in the central United States.  As at many similar institutions, our Classical Mechanics course is a textbook-centric 4 credit-hour course with a solid foundation in the basics of theoretical physics taught in a predominately lecture format. It is usually taken as the fourth physics course for physics majors and minors in the spring of their second year. Typically, the students are concurrently enrolled in a differential equations course.  Our course uses {\it Classical Mechanics} by J. R. Taylor \cite{Taylor}; the lectures cover the first 10 chapters, including topics such as Newton's laws of motion, momentum, angular momentum, energy, oscillations, Lagrange's equations, two-body central-force problems and rotational motion of rigid bodies. 

All homework problems were selected from the textbook and  were chosen to encourage students to practice solving problems. In return, students could earn extra points towards their final grade. Students were encouraged to work on problems in groups but, for pedagogical reasons, they were to write solutions independently.  Students in this class were not taught an explicit procedure to solve problems. Rather, they were free to solve them any way that they could, which gave us the opportunity to gain insight into their reasoning when left to their own devices.

In spring 2015, 12 students were enrolled in the class. They had 13 weekly homework assignments (each including 10-15 problems). A complete solution to each homework problem was expected to be about half a page long and include diagrams, mathematics and verbal statements. We scanned students' submitted homework for analysis before passing them to the grader. All students in the course consented to participate in our research study. 

\section{Analytic Framework}

Several attempts to model physics problem solving break it into discrete steps to be followed in a linear order~\cite{Heller92,Redish05}. However, unless students are explicitly taught to follow these algorithms, their approaches to problem solving rarely fit these linear models.  In contrast, studies using knowledge-in-pieces frameworks to understand students' non-linear problem solving ``in the wild'' (e.g., Ref.~\cite{Sayre08}) rely on the richness of interview or video data to make inferences about student reasoning.

The ACER framework bridges the gap between prescriptive problem solving and knowledge-in-pieces~\cite{Wilcox12, Wilcox13}. It was developed to analyze the use of mathematics in upper-division physics courses. The ACER framework is organized around four components: {\it Activation} (determining the proper mathematical tool); {\it Construction} (making a mathematical model); {\it Execution} (performing mathematical steps); and {\it Reflection} (checking the final solution). Within each of these four components, students may perform multiple specific actions that are then detailed using sub-codes.  For example, within Construction, a student might choose a coordinate system or set the limits of integration. In solving a particular problem, a student may freely move among these components in any order, visiting each of them several times with different sub-codes.  Naturally, the details of the sub-codes depend on the specific physics topic and problems at hand. 

Prior work using ACER framework has focused mostly on topics in electricity and magnetism. We extended the extant set of sub-codes to include codes appropriate for mechanics, as well as to cover students' errors (e.g., mistakes done when taking derivatives, using an incorrect strategy, etc.). We coded students' problem solutions, iteratively seeking new content codes and combining proposed codes until our code book covered $> 95\%$ of student work. As the code book approached stability, eight people participated in inter-rater reliability testing to assure the code book was a valid representation of student problem solving. When the code book was stable, multiple graders looked at different problems to confirm the reliability of coding.  Once two graders coded individual problems and achieved $> 90\%$ agreement on student work,  the code book was established. In this paper, we report on three problems selected from homework assignments across all students. We compare both across students and across problems to find patterns of mathematical tools use within each problem. 
	
To compare patterns in students' solutions, we used techniques borrowed from network analysis \cite{Kowalczyk}. A network graph is a collection of points, called nodes, and lines connecting these points, called edges. When the direction of a connection matters, we have a {\it directed graph}. In such cases, an edge from node A to B ($A \rightarrow B$) means something different than an edge from B to A ($B \rightarrow A$). If the direction of a connection does not contain any information about the relation between nodes (i.e, $A - B$ has the same meaning as $B - A$), a graph is called {\it undirected}. A graph can be characterized using various measures, such as node and edge count, diameter or density. Node and edge count correspond to the number of nodes and edges, respectively. Diameter is the longest graph distance between any two nodes in the network and it represents the linear size of a network. Density, on the other hand, is the ratio of the number of existing edges to the number of all possible edges, and can be thought of as a measure of network effectiveness. To determine the most important nodes in a network, one can perform centrality analysis. Depending on the nature of a network, as well as on the category of the ``importance of a node'' one is interested in, there are multiple centrality measures that can be used.\cite{Kowalczyk} 
Degree centrality indicates the importance of a node by the number of nodes connected to it, where the larger the degree, the more important the node is. 

Within this framework, nodes represent ACER sub-codes and the directed edges represent the order of steps students took to solve a given problem. When looking at the network created from all solutions to a given problem, we weight the edges of the graph by how many times students connected  a given two nodes to emphasize which connections between nodes are important. Edges with higher weights represent common connections within a network.   

\section{Variation in solutions within one problem\label{sec:net_dep_sin_stud}}
{\bf Problem 1}
A particle of mass $m$ is moving on a frictionless horizontal table and is attached to a massless string, whose other end passes through a hole in the table, where I'm holding it. Initially the particle is moving in a circle of radius $r_0$ with angular velocity $\omega_0$, but I now pull the string down through the hole until a length $r$ remains between the hole and the particle. (a.) What is the particle's angular velocity now? (b.) Assuming that I pull the string so slowly that we can approximate the particle's path by circle of slowly shrinking radius, calculate the work I did pulling the string. (c.) Compare your answer to part (b.) with the particle's gain in kinetic energy. 

\begin{figure}[t]
\centering
\includegraphics[scale=.52]{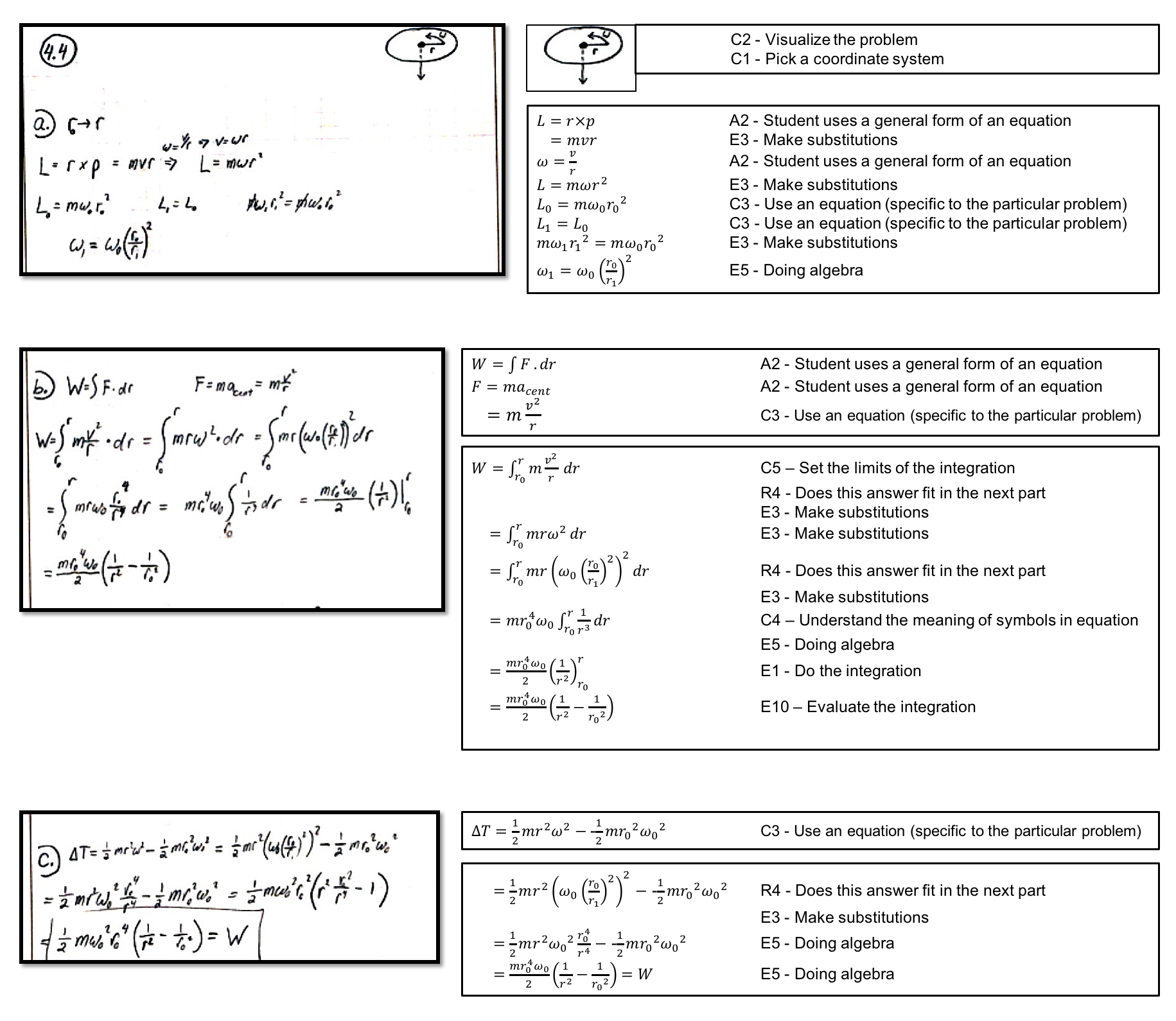}
\caption{\label{figure1}A solution to Problem 1 with a response coded according to the modified ACER framework.}
\end{figure}

\subsection{What was expected}
A solution for the sample problem is presented in \Fref{figure1}. This particular problem was not presented with a diagram. We expected that in order to visualize the physical system, the students would start with drawing one. In our ACER framework, that would be represented by codes {\it pick a coordinate system} (C1) and {\it visualize the problem} (C2). This problem has three parts and for all of them we expected students to begin with a general form of an equation, which would be coded as {\it use a general form of an equation} (A2). In part (b.), as they have to do an integration, we expected codes {\it set the limits of the integration} (C5), {\it do the integration} (E1) and {\it evaluate the integration} (E10) to occur. Both (b.) and (c.) require substitutions from the answer obtained in part (a.); therefore, we also expected codes {\it does this answer fit in the next part} (R4) and {\it make substitutions} (E3) to be used. 

As students perform mathematical operations in their solutions, we expect to see  {\it Execution} codes E3 and E5  ({\it make substitutions} and {\it do algebra}). When we consider the problem as a whole, we expect {\it Activation} code A2 ({\it use a general form of an equation}),  {\it Construction} codes C1, C2 and C5 ({\it pick a coordinate system}, {\it visualize the problem} and {\it set the limits of the integration}), {\it Execution} codes E1, E3, E5 and E10 ({\it do the integration}, {\it make substitutions}, {\it do algebra} and {\it evaluate the integration}, respectively) and {\it Reflection} code R4 ({\it does this answer fit in the next part}) to be dominating codes. 

\subsection{What we found}

When we used the ACER framework alone, we saw {\it Activation} code A2, {\it Construction} codes C1, C3 ({\it use an equation specific to the particular problem}) and C5, {\it Execution} codes E1, E3, E5 and E10 and {\it Reflection} code R4 as the most commonly used, dominant codes. Surprisingly, despite the lack of a figure in the problem statement, not all of the students started with visualizing the problem. They used both the general form of the equation and the problem related non-general form.

When we applied network analysis (NA) along with centrality measures to the ACER codes, we found that some of the nodes in the network, identified as important by the ACER framework (by the number of appearance for particular node), were not significant when looked at using the NA approach. On the other hand, network analysis identified nodes as more central which a frequency analysis using ACER alone would have overlooked. 

\begin{figure}[t]
\centering
\includegraphics[scale=.50]{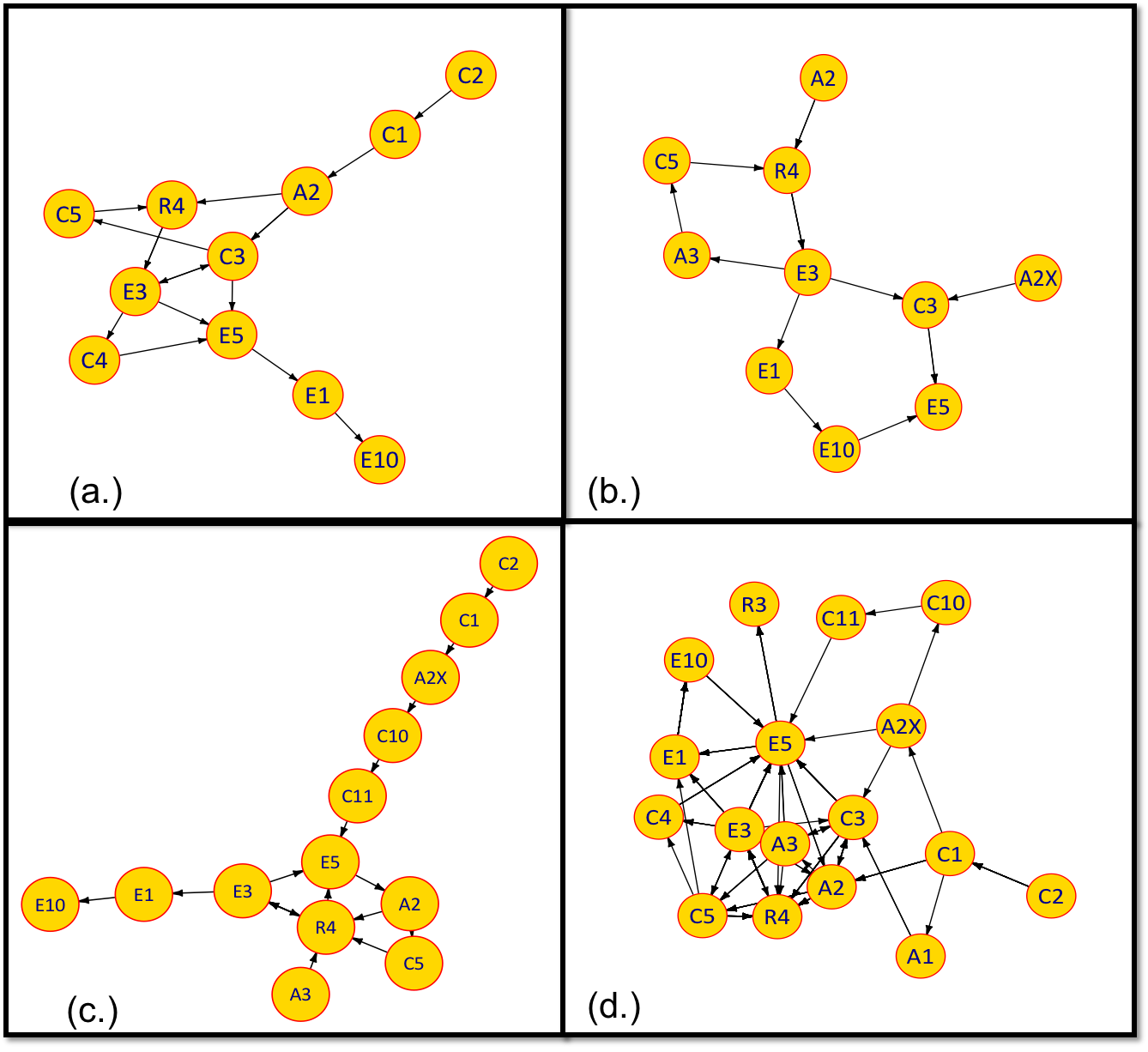}
\caption{\label{figure2}(a.) network graph representation of the solution presented in Figure~\ref{figure1}.,(b.), (c.) network graphs representing two other students' solutions to Problem 1 (students B, C),(d.) all students' responses to this problem merged into a single network }
\end{figure}

\Fref{figure1} shows a single student's solution coded according to the ACER framework. In \Fref{figure2}, the network graph for that student (Fig. 2a), two other students (Fig. 2b,2c), all students' responses combined into a single graph (Fig. 2d).

Even though this problem was a well-structured problem, different students had different approaches, as seen in the differences in their starting codes. As students in figure 2a.) and 2c.) took a diagram approach while student in figure 2b.) used equations at the start. The graph for all students (figure 2d) is denser than for any one student, and the specific patterns of each student are lost in the concatenation.

\Tref{tabl1} presents a summary of the basic network descriptives for all networks presented in \Fref{figure1}. When we look at sizes as measured by the diameter, we see the average diameter for the students was $5.82$ ($SD=1.59$). The average density of students' solutions was $0.16$ ($SD=0.05$) with values for individual student's ranging from $0.09$ to $0.23$.   

When we look at centrality indices for this problem, the node E3 ({\it making substitutions})  has the highest degree, indicating that it most frequently connects other steps in the problem solution. This is evidence that the mathematical operation of substitution is a crucial element in the solving process for this particular problem. Moreover, almost all the homework problems had multiple parts within the problem and often subsequent parts required a substitution found in a previous section. When we look at the network created by combining solutions to this problem from all students, we found that the top three nodes for all students are the same as the top nodes for both sample students (though their order is different). 

These network graphs show that students use two different approaches to solve this problem: they may start with visualizing the problem (such as with a diagram) before using equations in either a general or specific form; or they may omit the visualization and move directly to using equations. While this particular problem does not explicitly ask for graphical representation of the physical situation, drawing appropriate diagrams is a useful problem solving strategy \cite{Larkin87-WDM,Zhang94-RCT}, with the possibility that specific instruction on this topic may increase students' use of diagrams. 

\begin{table}[b]
\caption{\label{tabl1} Problem 1: Network characteristics for solutions from three students, for a network created by combining all students' solutions to this problem.}
\begin{indented}
\item[]
\begin{tabular}{@{}lcccc}
\br
  & Student A & Student B & Student C &All students\\
\mr
Diameter & 6 & 5 & 10 & 7  \\
Density & 0.155 & 0.167 & 0.115 & 0.632 \\ \br
\multirow{3}{*}{\pbox{5cm}{Nodes w/max. \\ degree}}& C3 and E3 & R4 and E3 & R4 & E3 \\
& E5 and R4 & C3 & E3 & R4\\
& A2 & E5 & E5 & E5\\ \br
\end{tabular}
\end{indented}
\end{table}

\section{The effects of problem statement type on students' solution graphs\label{sec:net_dep}}

We found that there is a dependence on the original problem statement reflected on the network graphs.  The first few problem solving steps that students use -- the first few nodes in the network -- depend on the kind of prompt that the problem uses.  In particular, we identified three different types of prompts that suggest to students how they should start the problem:

\begin{enumerate}
\item The problem statement is straightforward and asks students to perform specific mathematical operations, including trivial math procedures. An example statement looks like ``take the derivative/solve the above position equation for acceleration''. 

\item The prompt directs students towards the physical system or a diagram as the problem statement asks for an explanation of the physical system. An example prompt may include ``start with a diagram/free body diagram''. 

\item The problem statement directs students to think about what equations or conceptual resources\cite{Hammer2000} they should bring together to get an equivalent expression along with the physical system. To do so requires physics knowledge along with the correct mathematical steps. 
An example statement looks like ``show that/prove that\dots with the help of physical system''.\end{enumerate}

In order to determine whether students take the same approach when working on problems that are structured similarly, we identified three problems that belong to the same category (the second kind mentioned above) and we studied networks created by combining solutions to these three problems for each individual student as well as across all students.
 
\subsection{What was expected}
Since we compare problems structured similarly, we expect the crucial components of students' solutions to be similar across all students. Allowing for some slight variations, the core of the solutions should remain unchanged. The problems that we chose for our analysis include the one discussed in the previous section and thus we expected {\it Activation} code: A2 ({\it use a general form of an equation}), {\it Construction} codes: C1 ({\it pick a coordinate system}), C2 ({\it visualize the problem}), C3 ({\it use an equation specific to the particular problem}) and C5 ({\it set the limits of the integration}), {\it Execution} codes: E1 ({\it do the integration}), E3 ({\it make substitutions}), E5 ({\it do algebra}), E10 ({\it evaluate the integration}) and {\it Reflection} code: R4 ({\it does this answer fit in the next part}) to be the dominating codes.

\subsection{What we found}

\begin{table}[t]
\caption{\label{tabl2} Network characteristics for solutions from three students and for a network created by combining all students' solutions to three problems that were classified as structurally similar.}

\begin{indented}
\item[]
\begin{tabular}{@{}lcccc}
\br
  & Student C & Student D & Student E & All students \\
\mr
Diameter & 13 & 11 & 10 & 11 \\
Density & 0.133 & 0.277 & 0.211 & 0.309\\ \br
\multirow{3}{*}{\pbox{5cm}{Nodes w/max. \\ degree}}& E5 & E3 & E3 & E5 \\
& A2 & E5 & C5 & E3 \\
& C3 & C5 and A2 & E5 & C5 and C3 \\ \br
\end{tabular}
\end{indented}
\end{table}

Table~\ref{tabl2} shows characteristics of networks created by combining three structurally similar problems for three individual students, as well as of a network representing solutions from all students in the class combined. Nodes A2 ({\it use a general form of an equation}), C3 ({\it use an equation specific to the particular problem}), C5 ({\it set the limits of the integration}), E3 ({\it make substitutions}) and  E5 ({\it do algebra}) are most central for these networks.

As we expected based on the structure similarity, students used the general (A2) or non-general (C3) form of an equation to start the problem. Nodes E3 ({\it make substitutions}) and E5 ({\it do algebra}) again are key players in the network; however, since the problems include integrals, the node C5 ({\it set the limits of the integration}) becomes more prominent. By comparing networks across different problems, we can see whether students are consistent in their approaches to solving problems with similar structure. In all cases, node E5 ({\it do algebra}) was the most central node. Also, the four most prominent nodes (A2, C3, E3, E5) are the same for all students and for the merged network, although their order differs slightly. This alone could suggest that students take similar approaches to solve all three problems and that these approaches are quite similar among students.

\begin{figure}[t]
\centering
\includegraphics[scale=.55]{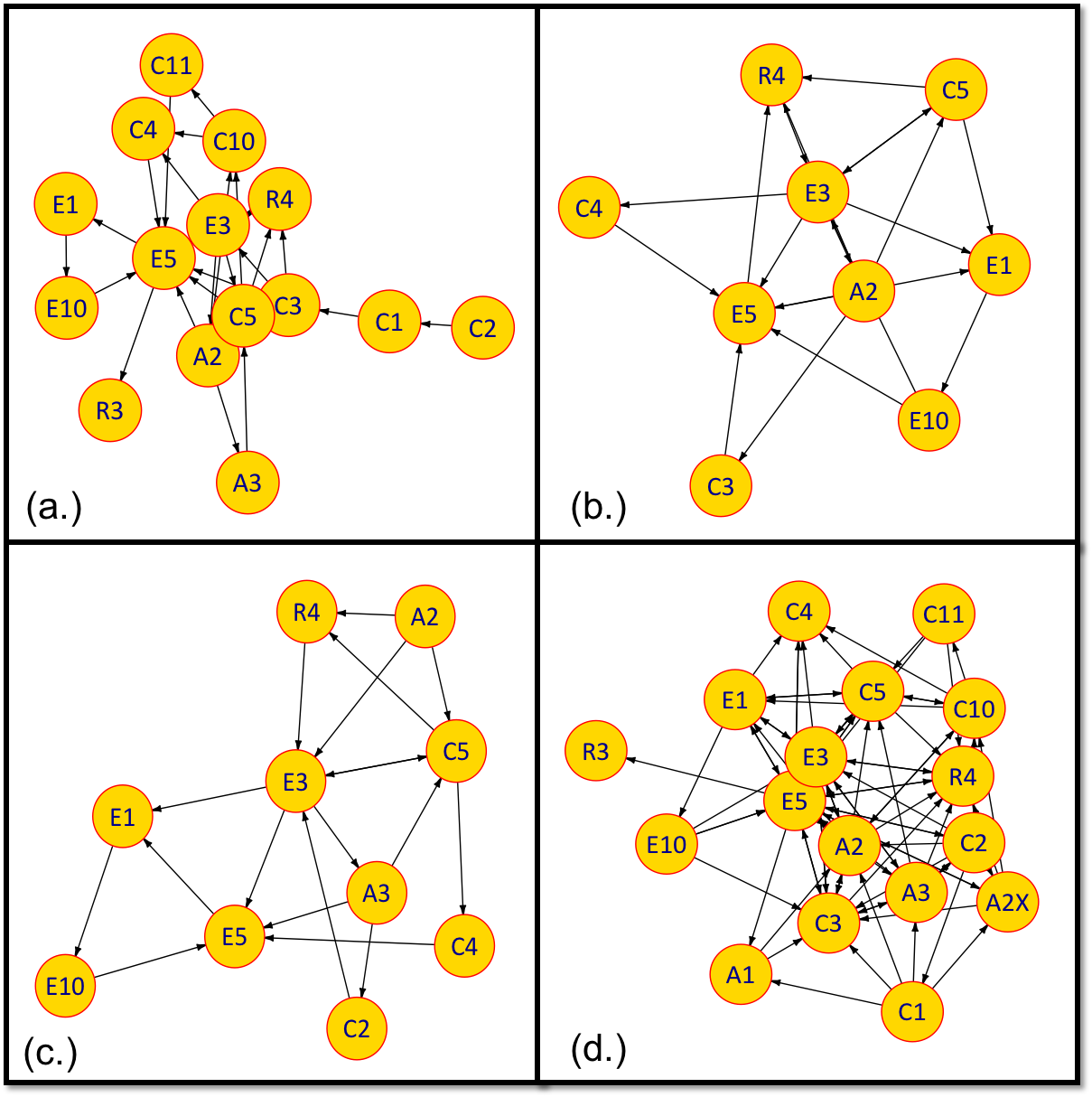}
\caption{\label{figure3}Network graphs created by combining solutions to three problems from three individual students (a. -- c.) and from all students' responses merged into a single network (d.).} 
\end{figure}

However, when we look at the network representing students' solutions in \Fref{figure3} we can see that the network presented in Fig. 3a is significantly different from the other two networks (Fig. 3b and 3c). Also, the network characteristics in \tref{tabl2} reveal that the network for student C has greater diameter and lower density (less connected graph with more nodes), indicating that this student got sidetracked in his solution. This could not be seen by the ACER framework alone, as the significant codes for this student were the same as for students who took a more direct approach.

\section{Problem statements and network complexity}

Further analysis suggests that problems consisting of more subsections lead to relatively more complex networks. Here we compare three problems that fall into different categories. Problem 2 belongs to the first type where students are required to perform trivial math procedures. Problem 3, on the other hand, falls into the second type and requires students to start with a diagram. 

{\bf Problem 2}
The shortest path between two points on a curved surface, such as the surface of a sphere, is called a geodesic. To find a geodesic, one has  to first set up an integral that gives the length of a path on the surface in question. This will always be similar to the integral 
\begin{equation}
L=\int_1^2\mathrm{d}s ={\int_x}_1^{x_2}\sqrt{1+{y'}^2(x)}\mathrm{d}s
\end{equation}
but may be more complicated (depending on the nature of the surface) and may involve different coordinates than $x$ and \textit{y}. To illustrate this, use spherical polar coordinates (\textit{r}, \textit{$\theta$}, \textit{$\phi$}) to show that the length of a path joining two points on a sphere of radius \textit{R}  is 
\begin{equation}
L=R \int_{x_1}^{x_2}\sqrt{1+{y'}^2(x)}\mathrm{d}s
\end{equation}

\begin{figure}[t]
 \centering
 \includegraphics[scale=.75]{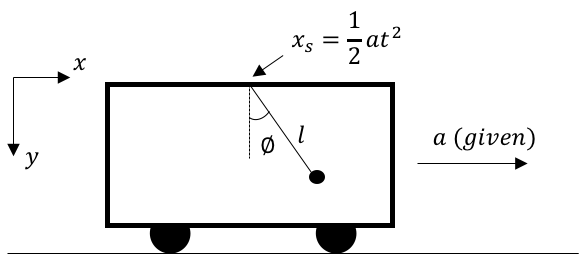}
 \caption{\label{figure4}Diagram for Problem 3.}
\end{figure}

{\bf Problem 3}
Consider the pendulum of Figure~\ref{figure4}, suspended inside a railroad car that is being forced to accelerate with a constant acceleration $a$. (a.) Write down the Lagrangian for the system and the equation of motion for the angle $\phi$. Use a trick similar to the one used in $x(t)=A\cos(\omega$t$-\delta)$ to write the combination of $\sin(\phi)$ and $\cos(\phi)$ as a multiple of $\sin(\phi+\beta)$ (b.) Find the equilibrium angle $\phi$ at which the pendulum can remain fixed (relative to the car) as the car accelerates. Use the equation of motion to show that this equilibrium is stable. What is the frequency of small oscillation about this equilibrium position?  

\subsection{The comparison}

In order to find the relation between the problem statement and the shape of the network here, we compare networks for three problems by all students. First, we compare the size of the networks and then we compare the most central nodes of each network. 

The size of a network is a measure of the number of nodes within a network. The network graph discussed in Section \ref{sec:net_dep_sin_stud} (Problem 1) has 16 nodes and 40 edges with eight strongly connected nodes. The network graph for Problem 2 has 14 nodes and 32 edges along with five strongly connected nodes and Problem 3 has a network graph with 11 nodes and 31 edges with two strongly connected nodes. The network graph of Problem 1 has a relatively large network size compared to the other two problems. This may be because Problem 1 has three subsections and the  other two problems have fewer subsections. 

\begin{figure}[b]
 \centering
 \includegraphics[scale=.4]{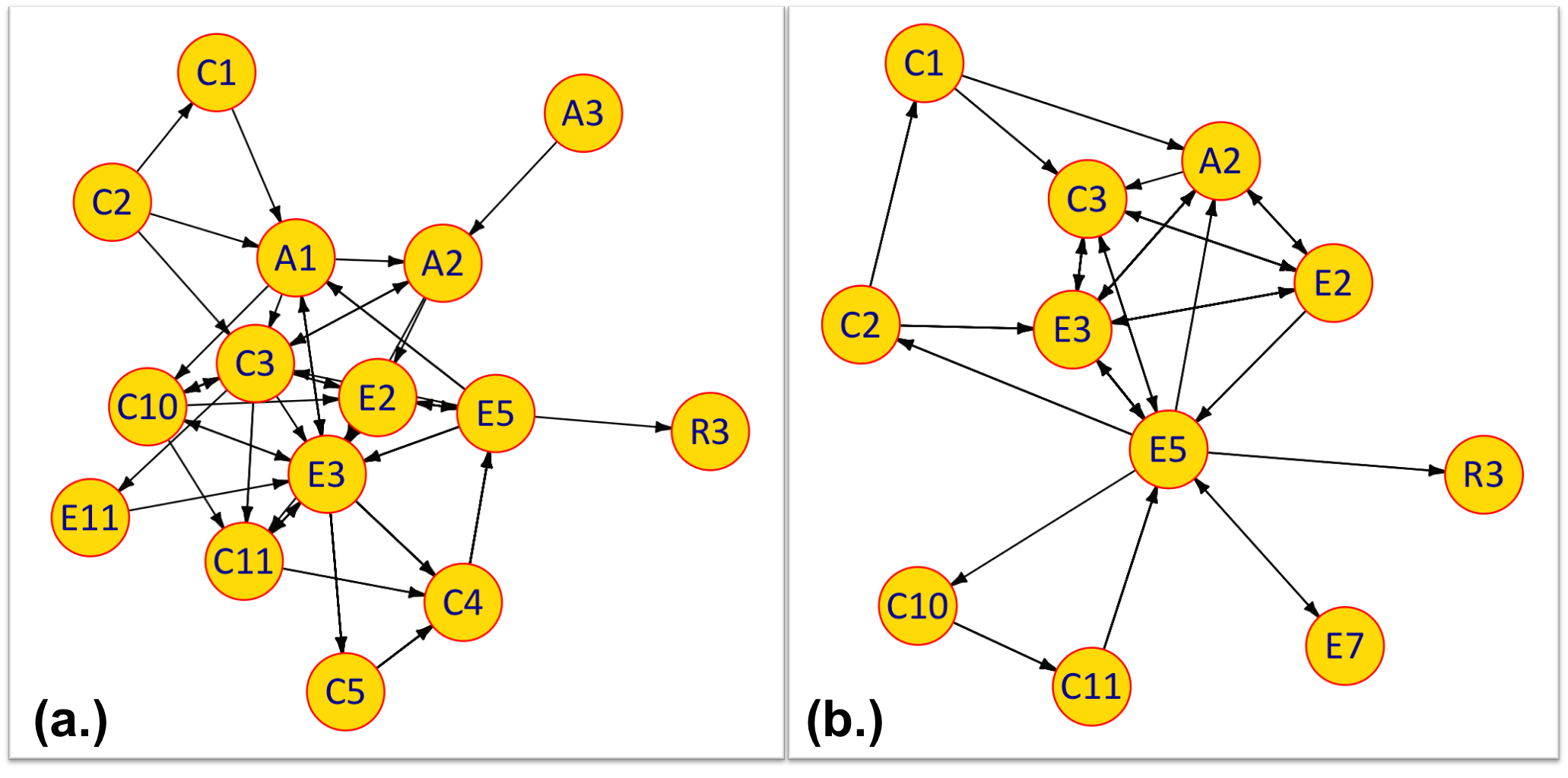}
 \caption{\label{figure5}(a.): Network graph for all responses to Problem 2. (b.): Network graph for all responses to Problem 3.}
\end{figure}

Network analysis shows that the most central nodes for the network graph for Problem 1 (\Fref{figure2}c -- all student responses) are E3 ({\it make substitutions}), R4 ({\it does this answer fit in the next part}), E5 ({\it do algebra}) and A2 ({\it use a general form of an equation}) with a network diameter of 5. According to the problem statement, students first have to understand the physical system and then use relevant equations to solve the problem. The network graph clearly shows that the students start with {\it Construction} codes, where they build a visual representation and then use the {\it Activation} codes, where they use appropriate equations. Because the problem has three subsections, students might recall an answer from a previous part and substitute it in the next part. Thus, R4 ({\it does this answer fit in the next part}) and E3 ({\it make substitutions}) are the most central nodes. 

E5 ({\it do algebra}), E3 ({\it make substitutions}), C5 ({\it set the limits of the integration}) and E2 ({\it take the derivatives}) are the most central nodes in the network graph for Problem 2 (\Fref{figure5}a -- all students' responses).  The network has a diameter of 5. The phrasing of Problem 2 also requires a visual representation.  The students start with {\it Construction} codes and with a diagram. Because the problem does not have subsections, R4 ({\it does this answer fit in the next part}) is not a central node. 

In contrast, E3 ({\it make substitutions}), E5 ({\it do algebra}), E2 ({\it take the derivatives}) and A2 ({\it use a general form of an equation}) are the central nodes in Problem 3 (\Fref{figure5}b -- all students' responses), with a network diameter of 4. Some different nodes appear in this network graph compared to previous two problems. E2 ({\it take the derivatives}) becomes a central node because students were asked to find the equation of motion and students take derivatives. Within their solutions, students use boundary conditions to show that this equilibrium is stable; this behavior shows up with two {\it Construction} codes: C10 ({\it state boundary conditions}) and C11 ({\it apply boundary conditions}). 

Comparison between problem statements and the network graphs shows that there is a dependence of problem statement on the shape of the network. The problems which consist of more subsections also lead to relatively complex networks.

\section{Limitations and Future work}

Our research question in this study was to see what, if anything, could be learned about students' problem solving given their homework solutions, as they represent a ubiquitous data source.   The information about students' performance revealed by the network graphs could be used to enhance the quality and consistency of the homework or quiz problems.  

However, coding homework solutions is a laborious process that is impractical for ordinary classroom use; instead these recommendations suggest that instructors compare students' solutions for elements in common (or unique to each student) as a pedagogical tool to gauge problem difficulty and divergent thinking. The NA approach is still more practical than interviewing and videotaping every single student, as it was done in the original studies~\cite{Wilcox12, Wilcox13}.

Another limitation of this data source is that we can't know if students go back and add more information later, because the only thing we have is the written solution. We coded students' solutions by assuming they wrote this strictly linearly; this is reasonable, as we have observational data of different students writing problems in mostly the order that they solve them in. In future work, we could more carefully test this assumption. 

Our data are drawn from one highly typical course at a single university.  This approach could be used to compare active-learning classrooms to traditional ones, or real-world problem sets to ones drawn from a textbook.   Though our data cover 23 questions overall, they are limited to classical mechanics.  To expand our study more broadly across the upper division, we are currently collecting data on an Electricity and Magnetism (EM) course, and augmenting students' written work with video recordings of them solving homework problems. 

	We could use a fine-grained ACER framework to focus on different parts of student solutions. For instance, we could use the fine-grained activation and construction to answer a research question like, what is the effect of different problem statements to students' approach to solve problems. To do this we may have to carefully design the problem statements. In our study we did not see large differences in students' approaches, perhaps because the textbook problems were not designed with that goal.

\section{Conclusion}
Network analysis has common applications in analyzing social interactions in groups of people, complex ecosystems and biological systems, and information transfer systems.  In this paper, we use network analytic tools to represent the relationships between the knowledge elements and steps of problem solving. We use the ACER framework to identify and code these elements and steps.  Together, network analysis and the ACER framework can model how students connect ideas to solve problems, and allow for quantitative comparisons among students and problems. Network analysis identifies the most important mathematical operations in the solving process, as interpreted through the ACER codes.

Network analysis indicates that in most of the analyzed cases, the most important nodes are E3 ({\it make substitutions}), E5 ({\it do algebra}), A2 ({\it use a general form of an equation}), R4 ({\it does this answer fit in the next part}). Substitution and algebra in general are crucial tools for solving these problems. Most of the homework problems had few parts within the problem and in these cases students may have to go back and check the previous solution and then substitute it in the next part.  While many problems have similar central nodes, the exact nodes and their order are different across students and across problems. The shape of the network graphs varies with different problem statements. 
Further research is needed to see exactly what properties of problem statements prompt students to use specific mathematical tools.  Nonetheless, network analysis of students' problem solutions reveals patterns in their tool use on typical problems.

	\section{Acknowledgments}

We thank Deepa Chari for her work with inter-rater reliability and the instructor and students for their participation. This work was funded by NSF DUE-1430967 and PHY-1344247 and the Department of Physics at Kansas State University.

\appendix 



\section*{Appendix - Codebook} 
\label{AppendixA} 

\begin{enumerate}
\item {\it Activation} of the tool 
	\begin{itemize}
  \item A1 - Identify the target (quantity/ value)
  \item A2 - Student uses a general form of an equation
  \item A3 - Student uses a less-general form of an equation
\end{itemize}

ANX (N=1, 2, 3)  used strategy is unhelpful or incorrect
\begin{itemize}
  \item A2X  Student uses an unhelpful /or other form of an equation
\end{itemize}

\item {\it Construction} of the model
\begin{itemize}
  \item C1 - Pick a coordinate system
  \item C2  Visualize the problem
  \item C3  use an equation specific to the particular problem 
  \item C4  understand the meaning of symbols in equation
  \item C5  set the limits of the integration
  \item C6  determine at which point the derivatives should evaluate
  \item C8  label forces on the free body diagram 
  \item C9  make assumptions
  \item C10  stating boundary conditions 
  \item C11  applying boundary conditions
\end{itemize}
CNX (N=1, 2,..7)  used strategy is unhelpful or incorrect

\item {\it Execution} of the mathematics  
\begin{itemize}
  \item E1    do the integration
  \item E2  take the derivatives
  \item E3  make the substitutions
  \item E4  student draw graph(s)
  \item E5  doing algebra
  \item E7  Approximations 
  \item E8  evaluate the derivatives
  \item C9  make assumptions
  \item E9 [previously C7]  take the cross product 
  \item E10  evaluate the integration
\end{itemize}
ENX (N=1, 2,..4)  used strategy is unhelpful or incorrect

\item {\it Reflection} on the result
\begin{itemize}
  \item R1  check the units
  \item R2  check the limits of the final answer
  \item R3  does this answer make sense? 
  \item R4  does this answer fit in the next part?
  \item R5  Comparing cases
\end{itemize}
RNX (N=1, 2 ...4)  used strategy is unhelpful or incorrect

\end{enumerate}


\section*{References}
\begin{thebibliography}{10}
\bibitem{Wallace10} C.S.~Wallace and S.V.~Chasteen, Upper-division students' difficulties with Ampere's law, {\it Phys. Rev. ST Phys. Educ. Res.} {\bf 6}, 020115 (2010).

\bibitem{Pepper12} R.E.~Pepper, S.V.~Chasteen, S.J.Pollock, and K.K.~Perkins, Observations on student difficulties with mathematics in upper-division electricity and magnetism, {\it Phys. Rev. ST Phys. Educ. Res.} {\bf 8,} 010111 (2012).


\bibitem{Wilcox12} B.R.~Wilcox, M.D.~Caballero, R.E.~Pepper, and S.J.~Pollock, Upper-division Student Understanding of Coulomb's Law: Difficulties with Continuous Charge Distributions, {\it Proceedings of the Physics Education Research Conference 2012} p. 1283, (2012).


\bibitem{Wilcox13} B.R.~Wilcox, M.D.~Caballero, D.A.~Rehn, and S.J.~Pollock, Analytic framework for students use of mathematics in upper-division physics, {\it Phys. Rev. ST Phys. Educ. Res.} {\bf 9}, 020119 (2013).

\bibitem{Kowalczyk} E.D.~Kolaczyk and G.~Csardi, Statistical Analysis of Network Data with R (Springer, 2014).

\bibitem{Taylor} J.R.~Taylor, Classical Mechanics (University Science Books, 2005).

\bibitem{Heller92} P. ~Heller, R. ~Keith, and S. ~Anderson, Teaching problem solving through cooperative grouping. Part 1: Group vs. individual problem solving, Am. J. Phys. 60, 627 (1992).

\bibitem{Redish05} E.~Redish, Problem solving and the use of math in physics courses, arXiv:physics/0608268 (2005).

\bibitem{Sayre08} E.C.~Sayre and M.C.~Wittmann, Plasticity of intermediate mechanics' students' coordinate system choice, {\it Phys. Rev. ST Phys. Educ. Res.} {\bf 4}, 020105 (2008).



\bibitem{Larkin87-WDM} J.~Larkin and H.~Simon, Why a Diagram is (Sometimes) Worth Ten Thousand Words, {\it Cognitive Science}, {\bf 11}, 65-69 (1987). 

\bibitem{Zhang94-RCT}J. Zhang and D. Norman, Representations in Distributed Cognitive Tasks, {\it Cognitive Science}, {\bf 18}, 87-122 (1994). 

\bibitem{Hammer2000} D. Hammer, Student resources for learning introductory physics. {\it American Journal of Physics, Physics Education Research Supplement}, {\bf 68} (S1), S52-S59 (2000).

\end{thebibliography}
\end{document}